\documentstyle[prd,preprint,tighten,aps,preprint]{revtex}

\begin{document}
\preprint{
\begin{tabular}{r}
JHU-TIPAC 950002
\\
CPHT-A345.0195
\end{tabular}
}
\title{Effects of the Running Gravitational Constant
on the Amount of Dark Matter}
\author{A. Bottino\thanks{E-mail Address, BOTTINO@TO.INFN.
IT},
C. W. Kim\thanks{E-mail Address, CWKIM@JHUVMS.HCF.JHU.EDU},
and J. Song\thanks{E-mail Address, JHSONG@ROWLAND.PHA.JHU.EDU}}
\address{*Dipartimento di Fisica Teorica, Universit\`a di Torino and\\
I.N.F.N., Sezione de Torino, via P. Giuria 1, 10125 Torino, Italy \\
and \\
\dag Centre de Physique Th\'eorique\\
Ecole Polytechnique, 91128 Palaiseau Cedex, France \\
and\\
\dag \ddag Department of Physics and Astronomy\\
The Johns Hopkins University\\
Baltimore, MD 21218, U.S.A.}
\maketitle

\begin{abstract}
\setlength{\baselineskip}{0.5cm}
The amount of dark matter in the Milky Way and beyond
is examined by taking into account the possible running of the
gravitational constant $G$ as a function of distance scale. If
the running of $G$, as suggested by the
Asymptotically-Free Higher-Derivative quantum gravity, is
incorporated
into the calculation of the total dark matter in the galactic halo,
the amount of dark matter that is
necessary to explain the rotation curve is shown to be reduced by
one third  compared with the standard calculations. However,
this running of $G$ alone cannot reproduce the observed flat behavior
of the rotation curve.
It is also shown that
the running of $G$ cannot explain away the presence of most of the
dark matter beyond the scale of $ \sim 10$ Mpc in the Universe.
We also present a pedagogical explanation for the running of $G(r)$
in the region of large scales which is clearly a classical domain.
\end{abstract}

\section{Introduction}
The nature (and the amount) of dark matter in our Milky Way and
the Universe
is one of the most important issues that we are facing in
particle physics and cosmology.
In spite of recent advances in establishing many limits on the
abundance of  various dark matter candidates and in estimating the
distribution of dark matter,  we are still far from understanding
the nature and the amount of dark matter. The recent observation of
the MACHOs (Massive Compact Halo Objects)[1] may shed some light
on the nature of dark matter in our Milky Way, when  statistics of
the observations further improves in the coming years.
Information about non-baryonic dark matter candidates will also
be provided in the future by the new low-background detectors for direct
detection and by other indirect means [2].

Recently, it
was shown that in the Asymptotically-Free Higher-Derivative
(AFHD) quantum gravity [3], the
gravitational constant, $G$,  may be asymptotically free.
Its consequences on the content of dark matter have been
briefly discussed in [4,5]. In this article we present
more detailed discussions on the amount of dark matter in our
Milky Way and in the Universe. Our discussions are based on
the recent work on the phenomenological consequences of the running
of $G(r)$ as a function of the scale $r$ for the cosmology described in
Ref.[6,7].
It is shown that if the $G(r)$ is assumed to be running
as indicated by the AFHD quantum gravity, the amount of dark matter
in our Milky Way is reduced by one third and the running of
$G(r)$
alone cannot explain the rotation curve. The behavior of $G(r)$ is
such that it can never
mimic all the dark matter distribution necessary to explain the
rotation curve in the Milky Way.

We also show that the increase of $G(r)$ as a function of $r$ is
not fast enough to explain the observed
variation of $\Omega_{0}(r)$  as
a function of distance scale. In particular, $\Omega_{0}(r)$ due to
the running of $G(r)$ becomes much slower than the observed
behavior beyond
the distance of $ \sim 10$ Mpc. Unless the running of $G(r)$ is
modified
drastically, it can never mimic most of the dark matter in the
Universe.

Finally we attempt to give a pedagogical explanation why the effects of
the running of $G(r)$ is sizeable only in a purely
classical domain where large distances are involved. The running of
the coupling constant in gauge theories is attributed to the quantum
effects. Therefore, it is expected that the effects are prominent
only in the quantum domain. In the case of the gravitational constant,
one anticipates the quantum domain to be in the region of the Planck
mass, $M_P$. The corresponding distance is the Planck length, $10^{-33}$
cm. The answer to this puzzle lies in the nature of asymptotic freedom of the
coupling
constant, which is inevitably subject to infrared slavery (i.e., confinement).

\section{Dark Matter in the Milky Way}

We begin with the well-known gravitational potential due to a
spherical
distribution of matter within radius $R$ with density distribution
$\rho(r)$ given by [8]
\begin{equation}
\Phi (r) = - 4 \pi G_{N} [ \frac {1}{r} \, \int _{0}^{r}dx\,x^2
f(x) +
\int_{r}^{R} dx\, x f(x)]~~,
\end{equation}
where we have used the definitions
\begin{equation}
G(r) \equiv G_{N}\, g(r)~~; ~~~~ G(0)= G_N~~,
\end{equation}
and
\begin{equation}
f(r) \equiv g(r)\, \rho(r)~~~.
\end{equation}
In Eqs.(1) and (2), $G_N$ is the Newton's gravitational constant.
It is to be noted that in Eq.(1), the standard  $\rho(r)$ was
replaced
by $f(r)= \rho(r)\,g(r)$ in order to take into account effects of
the running of $G(r)$, expressed in terms of $g(r)$. From Eq.(1),
the force is given by
\begin{equation}
| \vec{F}(r)| = 4{\pi} G_{N}\frac{1}{r^2}\int_{0}^{r}dx\,x^2
f(x)~~.
\end{equation}
When Eq.(4) is substituted into the equation of motion,
\begin{equation}
m |\vec{F}(r)|= m \frac{v^2(r)}{r}~~,
\end{equation}
one finds
\begin{equation}
v^2(r)=4 {\pi} G_N \frac{1}{r} \int_{0}^{r}  dx\,x^{2}f(x)~~.
\end{equation}
Taking derivative of Eq.(6) with respect to $r$ yields
\begin{equation}
f(r)= \frac{1}{4 {\pi} G_N} \frac{v^{2} + 2rvv'}{r^2}~~.
\end{equation}
It is customary to fit the observed rotation curve  by the following
two-parameter expression
\begin{equation}
f(r)= \frac{f(0)}{1 + [\frac{r}{r_c}]^2}~~,
\end{equation}
where $f(0)=\rho(0)$ and $r_c$ are, respectively, the galactic core
density and the size of the core. Comparison of Eqs.(7) and (8) at
large values of $r$ (with $v'=0$) gives the well-known expression
\begin{equation}
\frac{v^{2}(\infty)}{4 {\pi} G_N}=\rho(0)r^{2}_c~~.
\end{equation}
Hence, for a given $r_c$, $\rho(0)$ is determined by the observed
value $v(\infty) \simeq 220$ km/sec. However, this would
overestimate $\rho(0)$
because the presence of the spheroid is neglected in the above
discussion. The two standard derivations of $f(0)$ and $r_c$ are
due to Bahcall, Schmidt
 and Soneira [9] ,and Caldwell and Ostriker [10](see also [11,12]).
Here, we take, for definiteness, $r_c = 3$ Kpc and $ \rho(0)=
0.09 \, M_{\odot}pc^{-3}$ which includes the contribution of the
spheroid.

Now, if we adopt the $g(r)$ given in [6], with $r$ expressed in
units of
Kpc,
\begin{equation}
g(r) \equiv {1 + \delta_{G}(r)}= 1+ 0.3\,r^{0.15}~~,
\end{equation}
we obtain
\begin{equation}
\rho(r)= \frac{\rho(0)}{1+ [\frac{r}{r_c}]^2}\frac{1}{1+
\delta_{G}(r)}
= \frac{\rho_{S}(r)}{ 1 + \delta_{G}(r)}~~,
\end{equation}
where $\rho_{S}(r)$ denotes the standard density distribution
without the running $G(r)$. The behaviors of $4 \pi r^{2} \rho_{S}(r)$ and $4
\pi r^{2} \rho(r)$
are shown in Fig.(1), where the area under each curve represents the amount
 of dark matter in the halo for the corresponding case. The total amounts
of dark matter in the halo  of a radius 100 Kpc for the standard
(without $G(r)$) and modified (with $G(r)$) cases are given, respectively, by
\begin{equation}
M_{H}^{(S)}= 4 \pi \int _{0}^{100Kpc} \rho_{S}(x)x^{2} dx = 9.7 \times 10^{11}
M_{\odot}~~,
\end{equation}
and
\begin{equation}
M_{H} = 4 \pi \int _{0}^{100Kpc} \rho(x) x^{2} dx = 6.4 \times 10^{11}
M_{\odot}~~.
\end{equation}
Therefore, the running $G(r)$ reduces the dark matter content by one
third.
The amount of reduction is insensitive to the the size of the Milky
Way Galaxy. When integrated
up to $150$ Kpc in Eqs.(12) and (13), we have 36 $\%$ reduction of
the total dark matter content.

It is important to note that
what is needed to explain the rotation curve is the behavior of
$f(r)$,
 which is a product of $\rho(r)$ and $g(r)$, as given in Eq.(8).
Suppose we try to explain the dark matter content in the halo with
the spheroid and the running $G(r)$, then we would need $G(r)$ which is
increasing linearly in $r$ because the density of the spheroid goes as
$r^{-3}\sim r^{-3.5}$ for large distance. Since it is unlikely that $G(r)$
increases linearly in $r$,
it cannot mimic the dark matter in the halo. Therefore we conclude
that,
in spite of the running of $G(r)$, dark matter is necessary in the
halo,
 although the total amount can be reduced by one third.
Another related consequence is that the one third reduction of dark matter
in the halo would reduce the microlensing event rates by roughly the same
amount, ameliorating the apparent discrepancy between the standard
calculations of the event rates and the observed rates [13,14].
\section{Dark Matter in the Universe}
We now consider the amount of dark matter beyond the Milky Way.
In the earlier works [4,5] attempts were made to explain the dark
matter in
large scale structures by the increase of the running $G(r)$. The
$\Omega_{0}(r)$ was simply written as
\begin{equation}
\Omega_{0}(r)= \frac{8 \pi}{3} \frac{G(r) \rho_{0}}{H_{0}^2} \equiv
\overline{\Omega}_{0}
[ 1+ \delta_{G}(r)]~~,
\end{equation}
where $\rho_{0}$ and $H_{0}^2$ were assumed to be constant, i.e.
independent of
distance scale, and the local value $ \Omega _{0}(r \simeq 0)\equiv
\overline{\Omega}_{0}$ was taken to be $0.2$, which
is a very generous upper bound of $\overline{\Omega}_{0}$ obtained
from nucleosynthesis
arguments. Based on Eq.(14), the inferred dark matter content up
to the scales
of clusters ( up to the scale of the Virgo Cluster) was drastically
reduced
since  $\Omega_{0}$ given in Eq.(14) can mimic most of the dark
matter
except at very large scales beyond the Virgo cluster.

It was recently shown in [6,7], however,  that in a general scale-dependent
cosmology, the cosmological quantities
such as the Hubble constant and the age of the Universe
as well as the gravitational constant become all scale-dependent,
including  $\Omega_{0}(r)$ which is given by
\begin{equation}
{\Omega_{0}}^N(r) = \overline{\Omega}_{0} \frac {1 + \delta
(r)}{1 + \overline{\Omega}_{0} \, \delta (r)}~~.
\end{equation}
The difference between Eq.(14) and (15) is due to the fact that
in the standard Friedman cosmology, $\rho_{c,0}$ is taken to be a
constant, whereas it is $r$-dependent in this new scale-dependent
cosmology. Therefore, the behavior of ${\Omega_{0}}^N(r)$ in Eq.(15)
should not be
compared with the often quoted plot of the standard $\Omega_{0}(r)$
given in [15] which is essentially a quantity proportional to
the density itself. That is, in the new cosmology, $\overline{\Omega}
_{0}[1+\delta (r)]$, which is proportional to the density,
should be compared with the standard $\Omega_{0}(r)$ in
Eq.(14). Another important difference between them is that
$\Omega_{0}(r)$ in Eq.(14) continues to increase as $\delta_{G}(r)$
keeps increasing, whereas ${\Omega_{0}}^N(r)$ in Eq.(15) would reach unity
asymptotically and never exceeds unity.
Thus,
the ${\Omega_{0}}^N(r)$ in
Eq.(15) is a less rapidly growing function of $r$ than Eq.(14) for
the same $\delta(r)$ because of the denominator in Eq.(15).
In the following, we take the local value of $\overline{\Omega}
_{0}$ to be $0.1$ instead of $0.2$, which was used in [4], since the local
value determined in and
around
our Milky Way never exceeds $\overline{\Omega}_{0} \simeq 0.1$.

Now it is immediately clear from Fig.(2), where we have plotted
$\overline{\Omega}_{0}[1+\delta_{G}(r)]$ with Eq.(10), that, although
the running of $G$ can more or less mimic the observed matter density
within large horizontal and vertical error bars up to the scale of
$\sim 10 $ Mpc, it cannot explain  the dark matter beyond that scale.
In order to illustrate this  more specifically, we also plotted in Fig.(2) our
best fit of the
data points. (Admittedly, the best fit does not mean much
because of huge error bars but it was meant to guide the eye.)
This $best$ fit is given by
\begin{equation}
\delta(r)= 0.0028 \times r^{0.69}~~,
\end{equation}
where $r$ is given in units of Kpc.
Note that below $\sim 10$ Mpc, the contribution from the running of $G$
is larger than that of the $best$ fit. This  clearly cannot be the real
situation since
the observed fit should include all the effects from both the running of
$G$ and the dark matter. Note that
$\overline{\Omega}_{0}[1+\delta_{G}(r)]$
is slightly above the experimental error bars, indicating a possibility
that the running of $G$ in this region may be overestimated.
The important feature, however, is that beyond $\sim 10$ Mpc,
$\overline{\Omega}_{0}[1+ \delta(r)]$ completely takes over $\overline
{\Omega}_{0}[1+ \delta_{G}(r)]$ and becomes dominant. Although
Fig.(2) shows the crossover point at $\sim 10$ Mpc, it should serve as
a rough estimate. It is
not feasible, at present, to pin down the exact location of
this crossover point because
of the poor data quality. The crossover point depends very sensitively
on how one fits $\delta(r)$ with the very poorly known data. In Fig.(2),
the difference between the two curves beyond $\sim 10$
Mpc represents the true content of dark matter under the assumption that
the prediction on the behavior of $G$ given by Eq.(10) is correct.
In order to contrast the difference in another perspective, we compare,
in Fig.(3),  the behavior of the two ${\Omega_{0}}^N(r)$'s, one given by
Eq.(15) with
$\delta_{G}(r)$ in Eq.(10) and the other with
$\delta(r)$ in Eq.(16) which is nothing but the best fit-curve of
the data. Note that these curves are not
to be confused with the usual $\Omega_{0}(r)$ with a constant
$\rho_{c}$. Also, in this figure the difference between the two curves
represents
the portion of dark matter which cannot be explained by the
running of $G$.
( Because of the definition of
$\Omega_{0}(r)$, the areas under the curves do not represent the actual
amount of dark matter.)

So far we have discussed the difference between the previous treatment
of the running of $G$ as discussed in [4,5] and the one based on
the modified scale-dependent cosmology [6]. In the
latter, $\delta(r)$ was arbitrary to be determined phenomenologically.
In another version of scale-dependent cosmology [7]
a new metric and a generalized Einstein equation were introduced to
explain the same phenomenon. In this version,
the new metric dictates the form of $r$-dependence in $\delta(r)$ and
 the resulting crossover point appears at much larger scales.
The gap between the two $\overline{\Omega}_{0}[1+\delta(r)]$'s
in Fig.(2) widens very rapidly beyond the crossover point.
Future observations will decide which version is valid, if any.
It goes without saying that if and when the behavior of $G(r)$ is
modified from that of Eq.(10) due to future advances in better
understanding of quantum gravity, then the amount of dark matter as
discussed above
has to be modified accordingly.

\section{Discussions}
If we assume that the gravitational constant $G(r)$ runs as suggested
by the
AFHD quantum gravity, it is shown that the amount of dark matter
in the Milky Way necessary
to explain the observed rotation curve is reduced by one third
compared with the amount in the standard calculations with the
constant $G_N$. It is also argued that it is very unlikely that the
running of $G(r)$ can
mimic the total dark matter in the Milky Way.

The running of the newly defined $\Omega_{0}(r)$ in
the previous works [4,5], in which the running of $G(r)$ was added
in the standard Friedman cosmology in a straightforward manner, is
essentially proportional to the matter density. It is different from
the one defined in a new cosmology [6,7] with running cosmological
quantities, the difference being due to the fact that in this new
cosmology the critical density also increases as scale increases.
It was
pointed out that the correct local value of $\Omega_{0}$ should be
$0.1$
instead of $0.2$ which further reduces the previous [4,5] estimates
of
the contribution from the running $G(r)$. We have demonstrated that
the increase of the observed ${\Omega}_{0}$ as a function of $r$
is
much more rapid than that with the running of $G(r)$ alone, leading to the
necessity of a large amount of dark matter, in particular beyond
the scale of $ \sim 10$ Mpc.

Finally we comment on the running of $G(r)$. Some physics rationale
was given in [4] to justify the running of $G(r)$ in the classical
domain in spite of the common understanding that the running is due to
quantum effects. Here, we add another rationale which is purely
pedagogical.
The asymptotically-free running of $G(r)$ used in this article
was derived by using an effective AFHD quantum gravity,
which was motivated by supergravity, at one loop level.
The behavior of $G(r)$ is such that it stays as the Newton's
gravitational constant for scales up to around 1Kpc and then it
starts rising.
The obvious question is then `` Why are the quantum effects for the
gravitational constant so prominent in a purely classical region
where
large distances are involved ? ''  After all, the running of a
coupling constant according to the Renormalization Group Equation
(RGE) is the quantum effect in a gauge theory. Here, we
do not of course intend to reproduce the result of [4,5] but instead
we will illustrate why the running can be prominent in the classical
region.
The answer to this
question can be found in the behavior of the well-known example of the running
of the strong
coupling constant, $\alpha_{S}
(Q^2)$, which is also asymptotically free.
The one-loop expression for $\alpha_{S}(Q^2)$ is
\begin{equation}
\alpha_{S}(Q^2)=\frac{\alpha_{S}(\mu^2)}{1+
b\alpha_{S}(\mu^2)\ln{\frac{
Q^2}{\mu^2}}}~~,
\end{equation}
where $\mu^2$ is some fixed point and $b$ is given by
\begin{equation}
b=\frac{(11- {\frac{2}{3}}N_f)}{4\pi}~~,
\end{equation}
with the number of quark flavors, $N_f$.

Although Eq.(17) is the result of one-loop perturbative calculations,
which is valid beyond a certain value of  $Q^{2}$, we can conjecture,
because of infrared slavery or confinement, that $\alpha_{s}(Q^2)$ becomes
very large as $Q^2$ becomes smaller and smaller. This behavior is likely
to persist even to extremely small $Q^2$, which, when converted into
distance scale, corresponds to classical scales. Often, in order to
get an idea of the scale associated with the rising of $\alpha_{s}(Q^2)$,
one defines the value of $Q^2$ at which Eq.(17) diverges.
The $\alpha_{S}(Q^2)$
diverges
when $Q^2$ takes the value given by
\begin{equation}
Q^2 =\mu^{2}\exp {\lbrack{- \frac{1}{b\alpha_{S}(\mu^2)}}\rbrack
} \equiv \Lambda^2_{QCD}~~.
\end{equation}
Substituting, for example, $\alpha_{S}(\mu^2) = 0.3$ for $\mu^2 = 1 GeV^2$ into
Eq.(17),
we find
\begin{equation}
\Lambda_{QCD} \simeq 110~ MeV~~.
\end{equation}
This $\Lambda_{QCD}$ is the characteristic mass scale of the QCD.
The above property is inherent in the
asymptotically-free coupling constants in non-Abelian gauge
theories. (The fine structure constant, $\alpha$, does not have
this property,  the QED being an Abelian gauge theory.)

We
conjecture that
similar quantum effects are in operation for $G(r)$ at very large
scales.
In this case, the distances involved are larger than 1 Kpc. Hence, the
corresponding $\Lambda_{G}$ must be extremely small, say, $
10^{-35}
\sim 10^{-33}eV$, corresponding to  distances  of $10^{6}$ and
$10^{4}$
Mpc, respectively.
If $\Lambda_{QCD} \sim 110~ MeV$ were to be interpreted as
representing an effective gluon mass, then $\Lambda_{G} \sim
10^{-35}eV$ may be interpreted as an
effective graviton mass.
To be more specific, let us parametrize $G(Q^2)$ as, with the Plank
mass
$M_P$,
\begin{equation}
G(Q^2) \equiv \frac{ \alpha_{G} (Q^2)}{M^{2}_{P}}~~,
\end{equation}
where
\begin{equation}
\alpha _{G}(Q^2) = \frac{1}{1+ b \, \ln[\frac{Q^2}{M^{2}_{P}}]}~~.
\end{equation}
We caution the reader that the parametrization in Eq.(21) is, admittedly, not a
physically consistent one because it gives the
impression that
the gravity is generated by the exchange of a particle with mass,
$M_{P}$, as
 in the case of the effective weak-interaction coupling constant where the
force is generated by the exchange of the weak gauge bosons, $W$ and $Z$.
 We do not, of course, mean that. Rather,  we use the
parametrization in Eq.(21), based on dimensional arguments, to
facilitate our pedagogical discussion.

The $G(Q^2)$ defined above is asymptotically free for $b > 0$ and
becomes
identical to $G_N$ at $Q^2 = M^2_{P}$. Suppose that $b$ is very
small
(for which we have no rigorous explanation, but it is possible to have
very small $b$ if most of the contributions to $b$ from gauge bosons,
fermions and bosons, and their superpartners cancel with
each other ), then $G
(Q^2)$ remains the same as $G_N$ for the  usual particle-physics
range of $Q^2$. According to Eq.(22), $\alpha_{G}(Q^2)$ becomes
infinite when $Q^2$
takes the value given by
\begin{equation}
Q^2 =M ^2_{P} \, e ^{- \frac{1}{b}} \equiv \Lambda^{2}_{G}~~.
\end{equation}
It is now easy to see that if $b$ is very small, $\Lambda_G$
becomes very
small although the only mass scale involved is $M_P$. For $b=
\frac{1}{290}$, for example, we find $\Lambda_{G} = 10^{-35}$eV.
When $Q$ is converted into distance scale $r$, Eq.(21) with Eq.(22) and $b=
1/290$ yields qualitatively the same behavior of $G(r)$ obtained in
the AFHD quantum
gravity. This is our pedagogical explanation for the behavior of the
$G(r)$ at very large distance scales which are clearly a classical
domain.


This work was supported,
in part, by the National Science Foundation, U.S.A. and by the Research
Funds of the Ministero dell'Universit\`a e della Ricerca Scientifica e
Tecnologica.

\newpage
\begin{center}
\begin{large}
\begin{bf}
Figure Captions
\end{bf}
\end{large}
\vspace{5mm}\
\end{center}
\begin{description}
\item [Fig. 1] Profiles of $4\pi r^{2}\rho(r)$ as a function
of radius $r$ for the standard calculation without the running
$G(r)$ (dashed curve) and the one with $G(r)$ (solid curve).
\item [Fig. 2] Plots for $\overline{\Omega}_{0}[1+ \delta_{G}(r)]$
(dashed curve) and $\overline{\Omega}_{0}[ 1+ \delta(r)]$
which is a fit to data points (solid curve) as
functions of $r$. Data points represent ${\Omega}_{0}$ in the
standard definition with constant $\rho_c$.
Note the crossover point at $\sim 10$ Mpc, beyond which
$\overline{\Omega}_{0}[1 + \delta(r)]$ becomes dominant.

\item [Fig. 3] Plots for the fitted ${\Omega_{0}}^N(r)$ and
$\Omega_{0}^{N,G}(r)$ based on Eq.(15). Note that this
${\Omega_{0}}^N(r)$
is different from $\Omega_{0}(r)$ in Fig.2.
The shaded area indicates the portion of dark matter which cannot be
explained by the running of $G(r)$.

\end{description}
\end{document}